\documentclass[9pt,twocolumn,twoside]{pnas-new}

\templatetype{pnasresearcharticle} 
\usepackage{xcolor}

\title{Structural Origins of Cartilage Shear Mechanics}
\author[a,1,*]{Thomas Wyse Jackson}
\author[b,*]{Jonathan Michel} 
\author[b]{Pancy Lwin}
\author[c]{Lisa A. Fortier}
\author[b,$\dagger$]{Moumita Das}
\author[d,e,$\dagger$]{Lawrence J. Bonassar}
\author[a,$\dagger$]{Itai Cohen}

\affil[a]{Department of Physics, Cornell University, Ithaca, NY}
\affil[b]{School of Physics \& Astronomy, Rochester Institute of Technology, Rochester\, NY}
\affil[c]{Department of Clinical Sciences, Cornell University, Ithaca, NY}
\affil[d]{Meinig School of Biomedical Engineering, Cornell University, Ithaca, NY}
\affil[e]{Sibley School of Mechanical and Aerospace Engineering, Cornell University, Ithaca, NY}
\affil[*]{Equal first author contribution}
\affil[$\dagger$]{Equal corresponding author contribution}

\leadauthor{Wyse Jackson}

\authorcontributions{I.C., L.J.B., L.F., and M.D. designed research; T.W.J. performed the experimental research; T.W.J. analyzed data; J.M., P.L., and M.D. designed model and performed simulations; T.W.J., L.J.B., and I.C. wrote the paper with all authors contributing.}

\authordeclaration{The authors declare no conflict of interest.}
\correspondingauthor{\textsuperscript{1}To whom correspondence should be addressed. E-mail: tsw64@cornell.edu; ic64@cornell.edu}


\begin{abstract}
Cartilage, the soft tissue that coats the long bones in mammalian joints, is a remarkable material that is able to sustain millions of loading cycles over decades of use and outperforms any synthetic substitute. The bulk properties of this tissue reflect the mechanics of an extracellular matrix that is comprised of primarily two components: a collagen network and a reinforcing secondary network of aggrecan molecules linked by hyaluronic acid. Diseases of cartilage involve degradation of one or both of these networks either due to mechanical overloading or various biochemical processes. Crucially, how the loss of these constituents manifest in changes to the mechanical performance of the tissue, particularly in shear, is not sufficiently understood to enable predictive modeling of tissue performance in disease. 
Here, we present experiments and theory in support of a rigidity percolation framework that describes quantitatively the structural origins of cartilage's shear properties and how they arise from the mechanical interdependence of the collagen and aggrecan networks. This framework predicts that sensitivity to degradation is highly dependent on how close the collagen network is to its rigidity percolation threshold. Specifically, it explains that near the cartilage surface, where the collagen network is sparse and close to the rigidity threshold, slight changes in either collagen or aggrecan concentrations, common in early stages of cartilage disease, create dramatic changes in modulus that can lead to tissue collapse. 
More broadly, this rigidity percolation framework provides a map for understanding how changes in composition throughout the tissue alter its shear properties and ultimate in vivo function.

\end{abstract}

\significancestatement{
Healthy cartilage is a remarkable tissue, able to withstand tens of millions of loading cycles with minimal damage. While much is known about the structural origins of its compressive mechanics, how composition determines its mechanical behavior in shear, a major mode of failure, is poorly understood. Through a combination of microscale structure-function experiments and modeling, we develop a rigidity percolation framework to explain the structural origins of cartilage shear mechanics. This framework provides a new quantitative understanding for how the known degradative events in osteoarthritis determine the mechanical changes that are a hallmark of this disease.  
As such, this work provides a road map for understanding disease progression and in combination with non-invasive techniques such as MRI will enable more effective diagnosis and treatment.}

\dates{This manuscript was compiled on \today}
\doi{\url{www.pnas.org/cgi/doi/10.1073/pnas.XXXXXXXXXX}}

\begin{document}

\maketitle
\thispagestyle{firststyle}
\ifthenelse{\boolean{shortarticle}}{\ifthenelse{\boolean{singlecolumn}}{\abscontentformatted}{\abscontent}}{}


\dropcap{A}ffecting over 27 million people in the United States \cite{mora_knee_2018}, and over 250 million people worldwide, Osteoarthritis is one of the leading causes of disability. Osteoarthritis can arise from trauma, mechanical forces, inflammation, biochemical reactions, and metabolic changes to cells \cite{ayhan_intraarticular_2014,krishnan_cartilage_2018,shen_inflammation_2017}. As this disease progresses, inflammatory mediators can induce release of enzymes that result in degradation of the extracellular collagen and aggrecan networks \cite{berenbaum_osteoarthritis_2013}, the two most important constituents responsible for the mechanical properties of cartilage. The networks formed by these constituents are quite distinct in their properties. The collagen molecules form a network of fibrils with very high tensile strength. The aggrecan network is comprised of highly charged chondroitin sulfate molecules attached to a core protein in a bottle brush geometry. These aggrecan monomers combine with hyaluronic acid to form 10-1000 MDa aggregates that gel \cite{holmes_hyaluronic_1988}. The high amount of charge on the aggrecan network produces an osmotic stress that draws in water and swells the tissue. 
Broadly, damage to the collagen network leads to loss of tissue integrity \cite{krishnan_cartilage_2018} and a reduction in the capacity of the tissue to resist tensile \cite{williamson_tensile_2003} and shear strains \cite{kempson_tensile_1973,silverberg_structure-function_2014}. Loss of aggrecan reduces the osmotic swelling, makes the tissue more susceptible to compression \cite{bonassar_changes_1995}, and is also associated with the loss of shear properties \cite{palmer2009composition,wilson2007selective}. As these networks degrade, the tissue mechanics degrade as well \cite{setton_altered_1999, palmer2009composition,wilson2007selective}, though often in a non-intuitive, non-linear, and depth dependent manner, until eventually the tissue fails catastrophically. 
Importantly, understanding the path towards mechanical failure in cartilage requires knowledge of how the collagen and aggrecan networks contribute to function under compression and shear in both healthy and damaged tissue. 

A major step towards understanding how the collagen and aggrecan networks contribute to the compressive properties of cartilage were the developments of poroelastic \cite{frank1987cartilage} and mixture theories \cite{mow_biphasic_1980}, which account for water movement through the extracellular matrix, as well as additional theories that describe electrostatic/osmotic contributions to compressive mechanics \cite{buschmann1995molecular,basser1998mechanical}. 
These theories have been useful quantitative frameworks \cite{frank1987cartilage} to understand experimental observations describing how damage to the charged aggrecan network makes the cartilage more susceptible to compression. As the aggrecan network degrades, the osmotic stress driving tissue hydration dissipates and the drainage time-scale is significantly shortened due to larger effective pores in the extracellular matrix. This prediction is confirmed by studies that used enzymes such as chondroitinase abc and trypsin to model degradation of the aggrecan network and found that the bulk compressive modulus of a cartilage explant can decrease by up to 50$\%$ \cite{rieppo_structure-function_2003} and that the tissue hydraulic permeability can increase up to 15 fold \cite{bonassar_changes_1995}. Due to their quantitative predictive power, such theories have been important tools for understanding the compressive mechanics of healthy and damaged cartilage.

While these theories are effective at predicting the compressive behavior of the tissue, much less attention has been paid to the shear behavior and its dependence on tissue composition. Pioneering work has demonstrated that electrostatic contributions from the aggrecan network account for a significant portion of the shear modulus of healthy cartilage \cite{jin_effect_2001}. Additionally, it has been shown that the concentration-dependent shear properties of the collagen network are well described by a rigidity percolation model. Notably, however, a quantitative framework for the combined contributions of the collagen and aggrecan networks to the shear modulus of healthy and degraded cartilage does not exist. 
Because the shear mechanics of cartilage are of critical importance to its function in joints and because shear is the most common mode of failure for this tissue, developing such a framework to understand the shear mechanics of cartilage and its dependence on both the collagen and aggrecan constituents has been a major goal of the biomechanics community. 

\begin{figure}[ht!]
\centering
\includegraphics[width=0.95\linewidth]{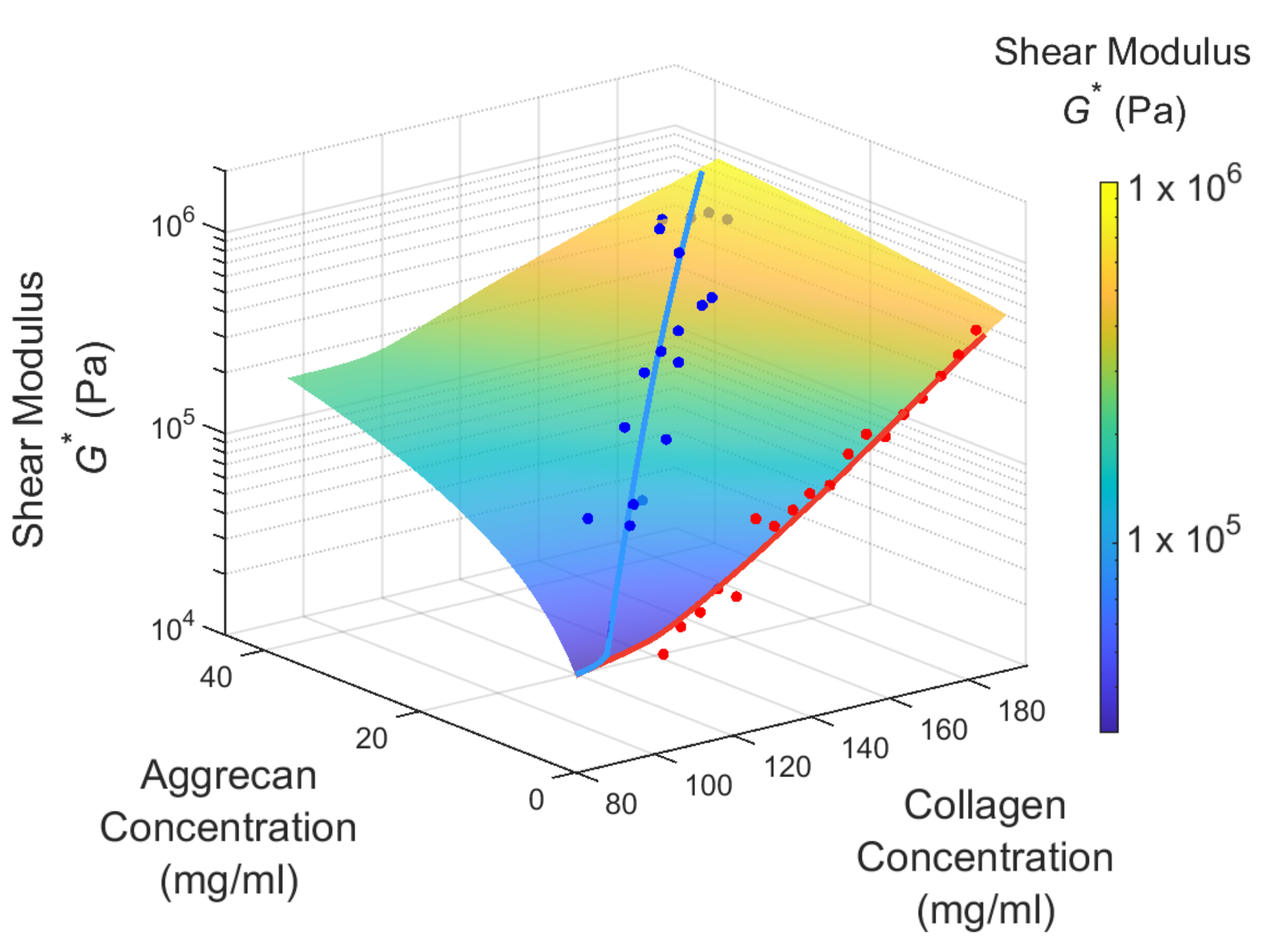}
\caption{{\bf Shear Modulus Dependence on Collagen and Aggrecan Concentrations: Experiments and Rigidity Percolation Framework Prediction.} The data points are experimental measurements of the modulus, measured via confocal elastography, as a function of collagen and aggrecan concentrations, measured via Fourier Transform Infrared Imaging. The surface depicts the theoretical prediction from the rigidity percolation model.}
\label{fig:3DMap}
\end{figure}

Here, we build on previous work \cite{silverberg_structure-function_2014} to develop a complete rigidity percolation framework for understanding the structural origins of cartilage shear mechanics (Fig.~\ref{fig:3DMap}). This framework reveals that describing the shear modulus of cartilage requires accounting for the critical interplay between the primary collagen and reinforcing aggrecan networks. It predicts that the shear modulus of cartilage is governed by how close the composite network is to the rigidity percolation threshold. Near this threshold, small compositional changes in either aggrecan or collagen drive large changes in shear mechanics, giving rise to a phase transition between a healthy, percolated, and mechanically competent network and a degraded, sparse, and mechanically incompetent network that can no longer sustain any shear loads. Such predictions yield particularly important insights for the mechanical behavior of the tissue near the cartilage surface where the collagen network is near the rigidity percolation threshold \cite{silverberg_structure-function_2014} and minor changes to the aggrecan network translate into large changes in the shear modulus \cite{griffin_effects_2014}. 

This paper combines novel experiments measuring local composition and shear mechanics with simulations of rigidity percolation phase transitions to develop and validate this model. Ultimately, this framework will enable predictions of how alterations in tissue structure and composition drive changes in mechanics that occur over the course of disease with the potential to inform diagnosis and therapy.

\section{Depth Dependent Composition and Mechanics}
\begin{figure*}[h!]
\centering
\includegraphics[width=1\linewidth]{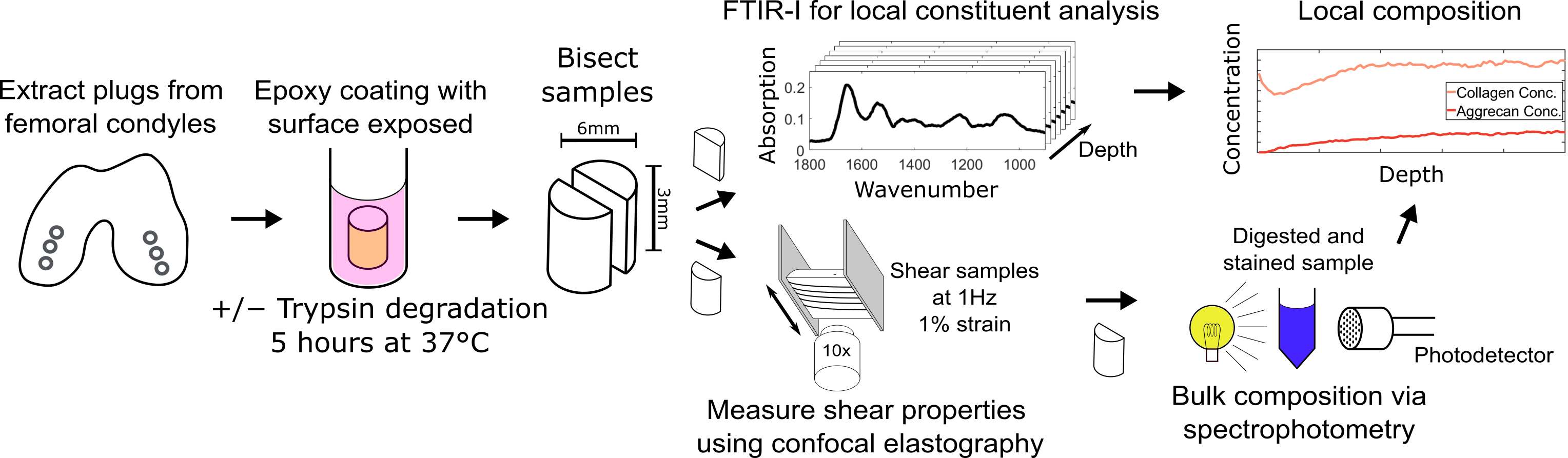}
\caption{{\bf Experimental Protocols.} Samples are extracted from medial and lateral condyles of neonatal bovid. A subset of the samples are covered in an epoxy coating leaving the surface exposed and submerged in a bath of trypsin for 5 hours at 37$^\circ$C. This subset constitutes the degraded samples. All samples are then bisected, with half being used for local compositional analysis with FTIR-I, and the other half being used for local mechanical analysis with confocal elastography. After mechanical testing samples are prepared for biochemical assays used to measure the absolute concentration of the constituents. These measurements are used to calibrate the FTIR data to determine the absolute concentrations of collagen and aggrecan.}
\label{fig:Method}
\end{figure*}

The complexity of the behavior predicted by the rigidity percolation model requires a large amount of data relating matched measurements of collagen concentration, aggrecan concentration, and shear modulus. To gather this data we apply techniques that we developed to measure and register composition and shear properties of cartilage on the microscale \cite{silverberg_structure-function_2014,middendorf_heterogeneous_2020,didomenico_measurement_2019}. Specifically, we use a combination of Fourier Transform Infrared Imaging and confocal elastography on matched samples of bovine cartilage as schematized in figure~\ref{fig:Method}. Importantly, the extracellular matrix of cartilage shows significant variation in collagen concentration primarily near the tissue surface \cite{silverberg_structure-function_2014}. By degrading the aggrecan molecules in this region we can obtain structure function relationships for the extracellular matrix that span concentrations in both the collagen and aggrecan networks. Towards this end, we harvested cartilage tissue plugs from femoral condyles and subjected a subset of the samples to tissue degradation with trypsin, which degrades the aggrecan core protein so that aggrecan leaves the tissue \cite{bonassar_changes_1995}. We then bisected all the tissue plugs. We measured the local tissue composition in one hemi-cylinder using histology and FTIR-I \cite{silverberg_structure-function_2014}. We measured the depth dependent shear modulus for the matching hemi-cylinder using confocal elastography \cite{buckley_mapping_2008,buckley_high-resolution_2010}. Using these measurements on healthy and degraded tissue we were able to quantify, the dependence of the modulus on a wide range of collagen and aggrecan compositions. 

\begin{figure*}[ht!]
\centering
\includegraphics[width=0.8\linewidth]{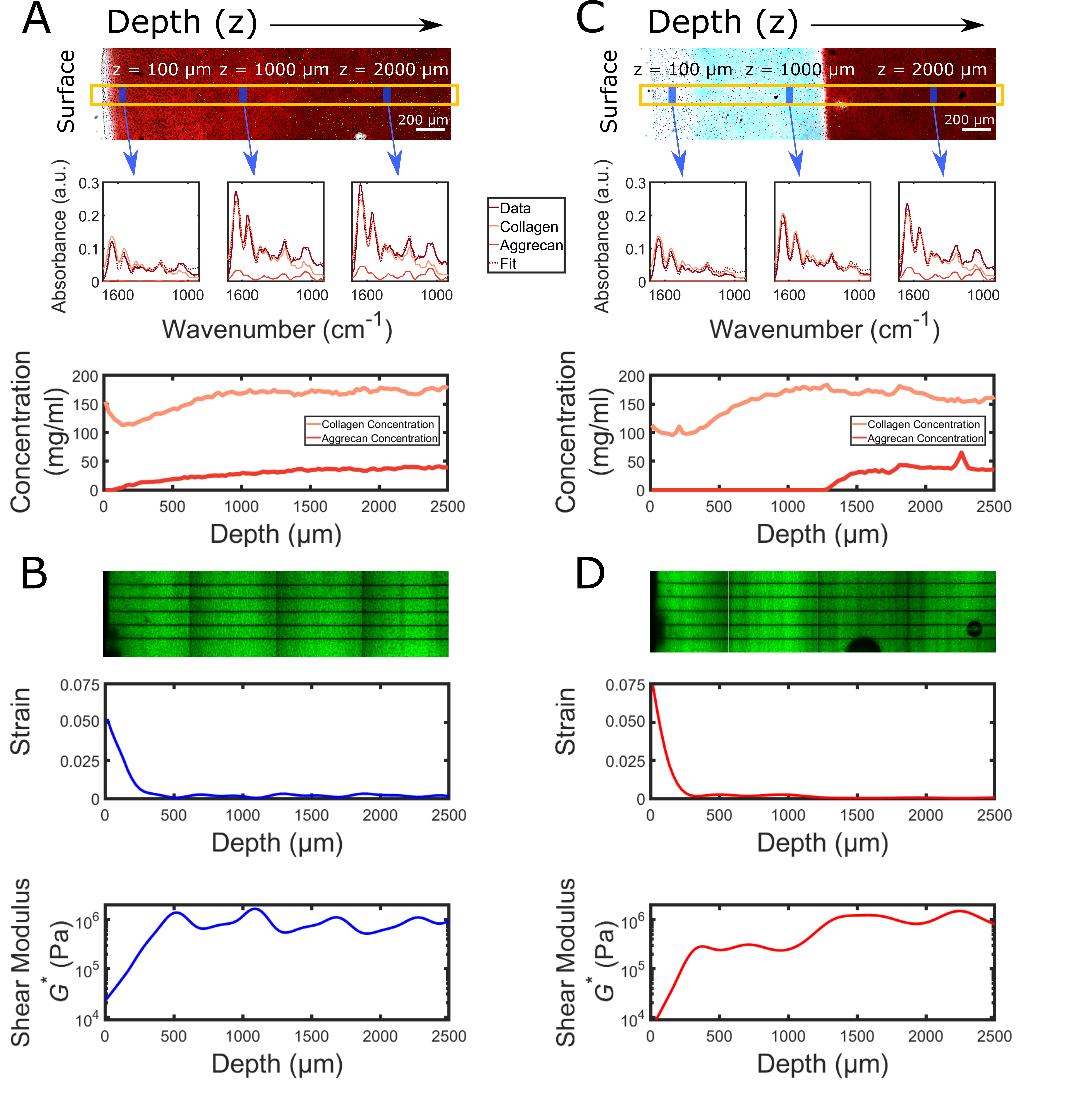}
\caption{{\bf Matched Structural and Mechanical Measurements.} A) Composition measurements. top: Sarafin-O stained histology slides for the healthy and degraded tissue. The areas stained red show regions of proteoglycan content. The yellow box indicates one of three sample spanning sections where the FTIR spectra were taken from, as well as three representative regions at depths $z$=100 \textmu m, $z$=1000 \textmu m, and $z$=2000 \textmu m for which the FTIR spectra are shown. middle: Measured absorbance spectra along with the fitted aggrecan and collagen contributions. The best fit spectra is also shown in dashed lines. bottom: Collagen and aggrecan concentrations versus depth.
B) Mechanical Measurements. top:
Lines are photobleached perpendicular to the tissue surface and shear oscillations are applied parallel to the surface (Methods). middle: By tracking the photobleached lines we extract the local strain within the tissue. bottom: These data are combined with measurements of the sample surface area and the total force needed to deform the tissue to extract the shear modulus as a function of depth. The same procedures are repeated for the degraded data to extract C) the collagen and aggrecan concentrations and D) the shear modulus for the degraded tissue as a function of depth.}
\label{fig:SingleSampleResults}
\end{figure*}

\subsection{Healthy Tissue}
Results of our depth dependent histology and Fourier Transform InfraRed Imaging (FTIR-I) measurements for healthy tissue are shown in figure~\ref{fig:SingleSampleResults}A. Safranin-O sections showed healthy cartilage morphology with increasing staining with tissue depth suggesting higher aggrecan concentrations deeper in the tissue (Methods). For each sample, the FTIR spectra as a function of depth was measured as described in the Methods section. Sample FTIR spectra at three different tissue depths ($z$=100 \textmu m, $z$=1000 \textmu m, and $z$=2000 \textmu m) are shown in the spectrograms in figure~\ref{fig:SingleSampleResults}A. Plotted are the absorption of infrared light as a function of wave number. The measured spectrum (red) is fit (dashed) by a sum of the contributions from the collagen spectrum (yellow), the aggrecan spectrum (orange), and a linear background (not shown) as described in the Methods section. We obtain excellent fits for all the spectra with the aggrecan spectrum contributing most notably to the peak at 1050 cm$^{-1}$. Consistent with the histology results, we find that the aggrecan contribution to the spectra is negligible at the tissue surface and increases with depth. 

From the best fit coefficients and whole-tissue biochemical assays performed on the samples (see Methods Section F for details of the assays), we determined the collagen and aggrecan concentrations with depth (bottom of Fig.~\ref{fig:SingleSampleResults}A). We find that the collagen concentration shows a typical dip just below the cartilage surface \cite{silverberg_structure-function_2014}. Specifically, we find that the concentration is 150 mg/ml at the surface, decreases by nearly 30$\%$ in the first 100 \textmu m and then increases again plateauing at 175 mg/ml beyond 800 \textmu m. We find that the aggrecan concentration monotonically increases from having a concentration of 0 mg/ml at the surface to a concentration of nearly 50 mg/ml in the deep zone. 

We used confocal elastography to determine the depth dependent shear modulus of the matching hemi-cylinder (Fig.~\ref{fig:SingleSampleResults}B). Briefly, we photobleached lines perpendicular to the articular surface and tracked their displacements with depth as described in the Methods section. The change in displacement over a given depth is used to determine the depth dependent strain. Using the extracted strain, surface cross section of the sample, and measurement of the total shear force, we determined the shear modulus $\it{G^*}$ (Methods). The measured modulus shows a typical response with a value of about 20 KPa at the surface that increases by almost two orders of magnitude over the first 500 \textmu m and remains roughly constant at greater depths \cite{buckley_mapping_2008,buckley_localization_2013,silverberg_anatomic_2013}. 


\subsection{Degraded Tissue}
Repeating this analysis on degraded tissue (Fig.~\ref{fig:SingleSampleResults}C), we find that the concentration profile of collagen remains unchanged while the aggrecan is completely removed throughout the degraded region. From histology, we observe that aggrecan removal proceeds as a front that is parallel to the tissue surface. The degraded region appears white or light blue while the region of tissue below the degradation front appears deep red. These results are consistent with previous histological measurements of trypsin degraded cartilage samples \cite{griffin_effects_2014}.   

For each sample, the FTIR spectra as a function of depth was measured as described in the Methods section. Sample FTIR spectra at three different tissue depths ($z$=100 \textmu m, $z$=1000 \textmu m, and $z$=2000 \textmu m) are shown in Fig.~\ref{fig:SingleSampleResults}C. Once again, we obtain excellent fits for all the spectra. Consistent with the histology results, we find that the aggrecan contribution to the spectra is negligible in the degraded region as indicated by the absence of the peak at 1050 cm$^{-1}$. From these data and the whole tissue biochemical assay we determined the collagen and aggrecan concentrations with depth. We find that the depth dependent collagen concentration remains similar to that of healthy tissue with a slight dip in concentration just below the surface and higher concentrations in the deeper regions. We find that the aggrecan concentration is close to zero in the degraded region and rises monotonically to levels similar to those in healthy tissues over a 200 \textmu m region at a depth of 1300 \textmu m.

As a result of this degradation we observe distinct changes to the depth dependent shear modulus of the matching hemi-cylinder (Fig.~\ref{fig:SingleSampleResults}D). We observe that the shear modulus at the surface is lower than in the healthy sample. For depths 400 \textmu m $< z <$ 1000 \textmu m we observe an intermediate value of the modulus that is about an order of magnitude larger than that at the surface and an order of magnitude lower than the modulus in the deep zone. The modulus in the deep zone is very similar to that of healthy tissue, because the degradation front never reaches this region.

\subsection{Average Depth Dependent Composition and Mechanics}
These trends in the tissue composition and mechanics for the healthy and degraded tissues hold when averaged across multiple samples (N=13 healthy, N=8 degraded). We plot the average collagen composition versus depth for both healthy and degraded samples in Fig.~\ref{fig:ConcentrationMap}A. We find no statistical difference between the healthy and degraded conditions. We plot the average aggrecan concentration versus depth for both healthy and degraded samples in Fig.~\ref{fig:ConcentrationMap}B. As in the single sample data in Fig.~\ref{fig:SingleSampleResults}, we find that aggrecan is completely removed from the fully degraded region for depths $z <$ 1200 \textmu m, indicating that our degradation protocol (exposure to trypsin at 0.25\% for 5 hours at 37 $^{\circ}$C) produces consistent results. Following this fully degraded region, we observe a transition region where the aggrecan rapidly increases in concentration until it reaches that of the healthy tissue and the two samples are statistically indistinguishable. We plot the shear moduli for the healthy and degraded samples in Fig.~\ref{fig:ConcentrationMap}C. We observe a consistent downward shift of roughly half an order of magnitude between the averages of the healthy and degraded samples. Moreover, the averages for the degraded samples continue to show an intermediate plateau region with a value between those of the surface and deep zones. While the modulus reduction is roughly constant across the degraded region, the difference in aggrecan concentration between the healthy and degraded tissues varies substantially. These data indicate that the dependence of the modulus on aggrecan varies with depth and depends on collagen concentration.   


\subsection{Structure Function Relationships}
\begin{figure*}[ht!]
\centering
\includegraphics[width=1.0\linewidth]{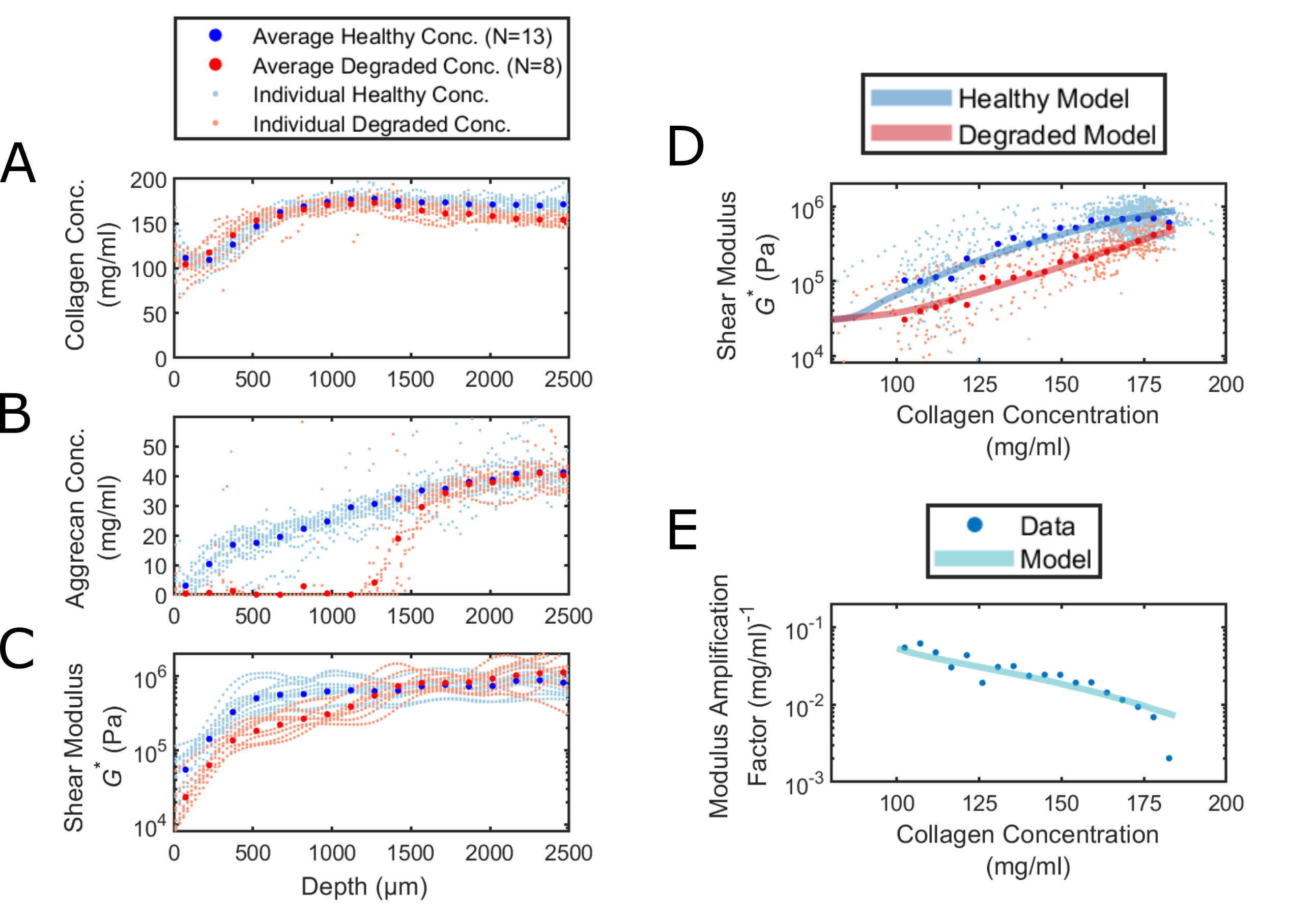}
\caption{{\bf Depth Dependent Results and and Shear Modulus Dependence on Collagen Concentration.} A) Collagen concentration with depth. Near the tissue surface there is significant variation in collagen concentration, until it plateaus after a depth of 1000 \textmu m. There is no significant difference between the healthy and degraded samples. B) Aggrecan concentration with depth. In the healthy samples, there is a natural increase in aggrecan concentration throughout the depth of the tissue. In the degraded samples, the aggrecan concentration is zero until the depth of the degradation front, where it returns to a similar concentration as the healthy samples. C) Shear Modulus with depth. The shear modulus at the surface of the degraded tissue is lower than in the healthy tissue. In the region 400 \textmu m $< z <$ 1000 \textmu m, there is an intermediate value of the modulus that is about an order of magnitude larger than that at the surface and half an order of magnitude lower than the modulus in the deep zone. The modulus in the deep zone is very similar to that of healthy tissue, because the degradation front never reaches this region. D) Shear modulus as a function of collagen concentration. The depth dependent measurements of the shear modulus, collagen concentration, and aggrecan concentration are combined to create a plot of shear modulus versus collagen concentration for both the healthy and the degraded tissue. The degraded tissue modulus is lower by close to a factor of 5. E) The modulus amplification factor as a function of collagen concentration: (Log$G^*_{Healthy}$ - Log$G^*_{Degraded}$)/($\Delta$ Aggrecan Concentration). The contribution of the aggrecan to the shear modulus is highest at low collagen concentrations and rapidly decreases with increasing collagen concentration.}
\label{fig:ConcentrationMap}
\end{figure*}

To determine the dependence of the modulus on both the aggrecan and collagen concentrations, we plot the local shear modulus as a function of collagen concentration for both the healthy (blue) and aggrecan degraded (red) tissues in Fig.~\ref{fig:ConcentrationMap}D. For the degraded tissue, we only include data with aggrecan concentration below 2~mg/ml since above this value the tissue has not been degraded and has the properties of healthy tissue. We find a uniform factor of 5 decrease in the modulus for the degraded tissue. 

Importantly, to ascertain the sensitivity of the modulus to aggrecan, we must account for the fact that the aggrecan concentration varies by almost an order of magnitude across these data sets. For example, at high collagen concentrations associated with the deep zone, reducing the aggrecan concentration from 30 mg/ml to zero reduces the modulus by a factor of 5. Remarkably, the same factor of 5 reduction in modulus results from reducing the aggrecan concentration from 3 mg/ml to zero near the surface region. To understand this dependence we define the aggrecan amplification factor as the difference in the Log of the healthy shear modulus minus Log of the degraded shear modulus divided by the change in aggrecan concentration. When the amplification factor is high, aggrecan plays a more important role in contributing to the shear modulus of the tissue. Conversely, when the amplification factor is low, the aggrecan concentration has a diminished role in determining the modulus. 

We plot the modulus amplification factor as a function of collagen concentrations in Fig.~\ref{fig:ConcentrationMap}E. We find that the contribution of aggrecan to the shear modulus is highest at low collagen concentrations and decreases by over an order of magnitude over a less than two fold change in collagen concentration. These measurements are consistent with a rigidity percolation framework to explain the shear mechanics of cartilage. At low collagen concentrations the network is less likely to percolate on its own and addition of aggrecan increases the likelihood of percolation and hence dramatically increases the shear modulus. At higher concentrations of collagen where the network is more likely to be percolated the presence of aggrecan makes a much smaller contribution to the shear modulus of the tissue. 

\subsection{Rigidity Percolation model}
\begin{figure}
\includegraphics[width=0.9\columnwidth]{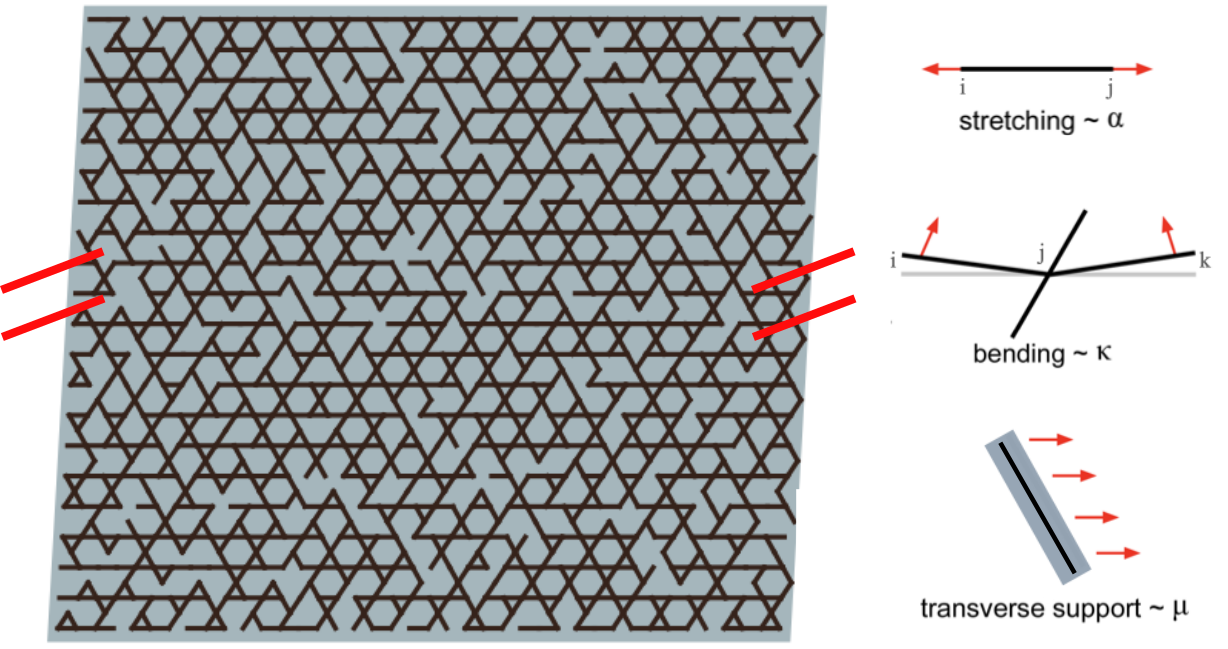}
\caption{{\bf Rigidity Percolation Model.} Shown is a schematic of a portion of the simulated kagome network. Hash marks indicate continuous boundary conditions in the lateral dimension. The black links represent the collagen fibers and the gray represents the background aggrecan gel. The links or bonds in the network are characterized by a stretching modulus $\alpha$ and a bending stiffness $\kappa$. The background gel further couples the links in the network through inhibition of transverse transport. This gel is characterized by a modulus $\mu$.}
\label{schematic}
\end{figure}

To quantitatively assess the degree to which our framework can describe both the healthy tissue and aggrecan depleted tissue moduli, we fit the data in Figs.~\ref{fig:ConcentrationMap}D,E to a rigidity percolation model. The model consists of a disordered kagome lattice representing the stiff primary collagen network embedded in a continuum elastic background gel representing the reinforcing aggrecan and hyaluronic acid networks (Fig.~\ref{schematic}). The fibers in the collagen network are randomly removed according to given probabilities $p$, where $0 <p <1$. Each bond is characterized by a stretching modulus $\alpha$ and a bending modulus $\kappa$. In a manner similar to \cite{silverberg_structure-function_2014} we include a background gel of aggrecan and other matrix components that resists bond deformations in the transverse direction and has an elastic modulus $\mu$. Here, however, to account for the fact that trypsin primarily affects mechanics through degradation of the aggrecan, we take $\mu$ to linearly depend on the concentrations of aggrecan, hyaluronic acid, and collagen (See Methods). The energy cost of deforming this composite network in the linear response regime is given by: 
\begin{eqnarray} 
E & = & 
\frac{\alpha}{2} \sum_{\langle ij \rangle} p_{ij}  \left (\textbf{u}_{ij} \cdot \hat{\textbf{r}}_{ij} \right)^2 \nonumber \\
&& + \frac{\kappa}{2} \sum_{\langle ijk \rangle} p_{ij} p_{jk}  \left[ \left( \textbf{u}_{ji} 
+ \textbf{u}_{jk} \right) \times \hat{\textbf{r}}_{ji}  \right]^2  \\ \nonumber
& & + \frac{\mu}{2} \sum_{\langle ij \rangle}   p_{ij} \left[ \textbf{u}_{ij}^2 -  \left( \textbf{u}_{ij} \cdot \hat{\textbf{r}}_{ij} \right)^2 \right] + \frac{\mu}{2} \epsilon_s^2 A,
\label{sim1}
\end{eqnarray}
where the terms correspond to the energy penalty for fiber stretching, fiber bending, the coupling of the network to the background gel, and the deformation of the background gel, respectively. In the final term, $\epsilon_s$ denotes an affine background shear strain, while $A$ is the area of the network. The indices $i,j,k$ refer to sites (nodes) in the lattice-based network, such that $p_{ij}$ is 1 or 0 when a bond between those lattice sites is or is not present. The quantities $\hat{\textbf{r}}_{ij}$ and $\textbf{u}_{ij} = \textbf{u}_{i}-\textbf{u}_{j}$ are respectively the unit vector along a bond $ij$ and the corresponding relative displacement. 

To determine the network properties we strain the network, use conjugate gradient minimization of the deformation energy (Eq.~\ref{sim1}) to mechanically equilibrate the network, and extract the shear modulus (See Methods). Specifically, for each set of parameters ($\alpha$,$\kappa$,$\mu$), a network containing $\sim 10^5$ nodes was randomly generated with a fraction $1-p$ of bonds missing, subjected to a compressive strain of $5\%$ and a shear strain of $1\%$ applied via the top boundary, and then allowed to relax via fiber deformations, with periodic boundary conditions imposed along the left and right sides of the network. With these simulations we obtained the modulus as a function of the bond occupation probability $p$ and the gel modulus $\mu$.




To map the results of the simulation to experiments, we linearly scale the network modulus
, bond occupation probability $p$, and gel modulus $\mu$ to the experimentally measured modulus, collagen concentration, and aggrecan concentration, respectively. We then fit the model to the collagen dependent modulus and amplification factor data in Figs.~\ref{fig:ConcentrationMap}D and \ref{fig:ConcentrationMap}E and find excellent agreement. In all, the rigidity percolation predictions for both the healthy and degraded data sets are obtained using 5 fitting parameters. Moreover, the fits for the bending to stretching ratio $\kappa/\alpha$ and the modulus of the bare hyaluronic acid gel are in very close agreement with literature values \cite{poole_immunoelectron_1982,yadavalli_nanoscale_2010,mow1992cartilage,jin_effect_2001}. Importantly, once these parameters are fit to the data, it is possible to use this rigidity percolation framework to quantitatively predict the modulus for arbitrary values of the collagen and aggrecan concentrations as shown in Fig.~\ref{fig:3DMap}.

\section{Discussion}
By collecting a large amount of matched experimental data for local composition and local shear mechanics and fitting them using simulations, we have constructed a rigidity percolation framework that provides valuable insights into the dependence of cartilage
mechanical properties on the tissue constituents (Fig.~\ref{fig:3DMap}). 
Our measurements and simulations show that the contribution of aggrecan to the shear modulus is highly dependent on the concentration of collagen. When the collagen concentration is high, the aggrecan provides a relatively minuscule contribution to the modulus. Conversely, aggrecan plays a critical role in enhancing the shear properties of cartilage in regions of the extracellular matrix where collagen concentration is low (Fig.~\ref{fig:ConcentrationMap}E). By capturing the interactions between the supporting aggrecan gel and the collagen network, the rigidity percolation framework provides important intuition for the origins of this non-linear and unusual behavior. When the collagen network is sufficiently concentrated that it percolates by itself, the relative contribution of aggrecan is small. Conversely, when the collagen network is close to the rigidity percolation threshold, even a small reinforcement of the network by aggrecan makes it easier for the composite network to transmit stresses. In combination, the model and experimental data provide a powerful framework for understanding how the shear mechanics of cartilage arise from the interactions of the collagen and aggrecan networks. 

This rigidity percolation framework for describing how 
collagen and aggrecan interact to determine the shear mechanics of articular cartilage is remarkably effective despite its simplicity. The model uses a conventional 2D Kagome lattice structure to represent the crosslinked collagen network that is coupled to a background reinforcing network representing the contributions of aggrecan and other matrix components. The contributions of the collagen network concentration, connectivity, and crosslinking, are effectively described by the bond occupation probability, $p$, and the lattice structure. Future studies using different enzymes that target these structural aspects of the collagen network would be helpful in elucidating their effective relative contributions to $p$. 

Importantly, a critical contribution of the reinforcing background gel is to provide a coupling between the fibers in the network. As such, its contribution is not only additive as has been suggested in prior literature \cite{jin_effect_2001}. As shown in the supplementary information (SI), when we exclude the third term in Eq.~\ref{sim1}, which couples the network deformations to the background gel, we obtain inferior fits and unreasonable values for the fitting parameters. These include a bending to stretching ratio that approaches 1 and a value for the hyaluronic acid gel contribution to the modulus $\sim 1$ Pa, which is three orders of magnitude too small \cite{holmes_hyaluronic_1988}. These findings highlight that the aggrecan gel can dramatically alter the shear properties of cartilage by helping to drive the composite network through its rigidity percolation transition. This mechanism for controlling the tissue properties is important biologically since the turnover rate for aggrecan is orders of magnitude faster than the turnover rate for collagen. As such, fabrication and degradation of aggrecan by chondrocytes, the cells in cartilage, can be used to rapidly alter the tissue shear mechanics in response to changes in shear loads.

  

A model that describes the dependence of cartilage shear mechanics on composition is also a powerful tool for understanding the progression of diseases of cartilage such as arthritis. Such diseases develop over many years and are typically characterized by slow degradation of one or more components of the extracellular matrix, but at late stages often result in rapid loss of joint function due to compromised cartilage mechanics. In the context of the proposed framework, this rapid loss of function may arise from the rapid decrease in mechanical integrity associated with crossing the rigidity percolation threshold. Recent advances in clinical MRI of cartilage tissue enable mapping of collagen and aggrecan composition and organization in vivo \cite{burstein2000mri}. Relating such measurements to tissue and joint scale mechanics would be a major advance in the field. The rigidity percolation framework presented here provides an opportunity to accomplish this important goal and may even allow for identifying tipping points during disease progression where small additional changes in composition lead to tissue failure. This understanding, could be critical for informing treatment by identifying stages of disease progression that are most in danger of compromising function due to loss of collagen, aggrecan, or both. 

Additionally, this framework provides new insights for understanding and designing cartilage therapies. Specifically, regenerative medicine approaches frequently involve delivery of cells alone \cite{brittberg1994treatment} or in combination with weak scaffolds \cite{griffin2015mechanical,van2019emergence} to promote cartilage regeneration. Critical to the success of such approaches is the generation of a new mechanically competent extracellular matrix. Cells alone or those embedded in sparse matrices are likely far below the rigidity percolation threshold and, as such, will take a significant amount of time to achieve mechanical competence. The framework described here suggests that designing implants close to the rigidity percolation threshold will maximize the impact of cellular matrix biosynthesis on their mechanical performance.

\section{Conclusions}
In this work we developed a rigidity percolation framework to understand the structural origins of cartilage shear mechanics based on the composition and interactions of the collagen network and the reinforcing aggrecan gel that together account for the majority of the tissue extracellular matrix properties. Since these are also the main constituents of all extracellular matrices in connective tissue in mammals, this framework and its extensions into nonlinear deformation regimes may be a widely applicable tool for understanding the mechanics of many if not all connective tissues in health, disease, and repair \cite{van2019emergence}. Similarly, this framework may also prove useful for understanding artificial constructs with tissue-like properties \cite{van2019emergence}. More broadly, this work illustrates a striking example of how biology exploits compositional perturbations driving physical processes at the proximity of phase transitions to achieve remarkable function in tissue homeostasis, disease, and repair.

\matmethods{\subsection{Tissue Harvesting}
One- to three-day old neonatal bovine knee joints were acquired from a local abattoir (Gold Medal Packing, Rome, NY). Neonatal cartilage was chosen due to its similarity in shear modulus profile to human cartilage \cite{Temple_Viscoelastic_2016, buckley_high-resolution_2010}, as well as ease of access, consistency of tissue samples, and prior use in similar studies \cite{buckley_mapping_2008,buckley_high-resolution_2010,buckley_localization_2013,silverberg_anatomic_2013,griffin_effects_2014,silverberg_structure-function_2014}. Cylindrical explants 3 mm in height and 6 mm in diameter were dissected from the medial and lateral femoral condyles.

\subsection{Enzymatic Degradation} Samples were enzymatically degraded by immersing them in a bath of trypsin-EDTA 0.25\% (Sigma) at 37°C for five hours (Fig.~\ref{fig:Method}). Epoxy glue was used to create a protective coating around the sample edges, with the top surface of the tissue left exposed. This procedure allowed for creating a well-controlled degradation front. Trypsin was chosen for its ability to primarily affect mechanics through degradation of the aggrecan and minimal effect on the collagen network \cite{liu_action_2018}. Post degradation, the samples were rinsed with PBS, the glue was peeled off, and the samples were bisected into two hemi-cylinders. One hemi-cylinder was placed in protease inhibitors for mechanical testing, and the other was fixed in 10\% PBS-buffered formalin for compositional analysis via Fourier Transform InfraRed Imaging (FTIR-I).

\subsection{Histology} Qualitative analysis of aggrecan removal was conducted via Safranin-O staining as described previously \cite{griffin_effects_2014}. Tissue samples were fixed in neutral buffered formalin, embedded in paraffin, cut into 4 \textmu m thick sections, and  placed onto glass slides. Sections were dewaxed in three xlyene baths for 2 min each and rehydrated in three baths of ethyl alcohol (100, 95, and 70\% ethanol, appropriately diluted with distilled water) for 2 min each. The nuclei were stained with Weigert's Iron hematoxylin for 10 min, then the samples were rinsed for 10 min in running water. The samples were then stained with fast green solution for 5 min, and rinsed with 1\% acetic acid solution for 15 seconds. Finally the samples were stained with 0.1\% Safranin-O, pH 2.0, solution for 8 min.

\subsection{Compositional Measurements} Quantitative measurements of collagen and aggrecan relative concentrations were obtained via Fourier Transform InfraRed Imaging similarly to previous studies \cite{Potter_Imaging_2001,camacho_ftir_2001,silverberg_structure-function_2014,middendorf_heterogeneous_2020,didomenico_measurement_2019}. Sections, 4 \textmu m thick from each tissue sample, were placed on 2-mm-thick mid-infrared (IR) transparent BaF$_{2}$ disks that were 25 mm in diameter (Spectral Systems, Hopewell Junction, NY). Although the sectioning plane was fixed, the section cutting direction was randomized to prevent systematic biases in section thickness due to cutting. Sections were dewaxed in three xylene baths for 2 min each and rehydrated in three baths of ethyl alcohol (100, 95, and 70\% ethanol, appropriately diluted with distilled water) for 2 min each.

Samples were loaded into a Hyperion 2000 Fourier Transform InfraRed Imaging (FTIR-I) microscope (Bruker, Billerica, MA) in transmission mode and we acquired data for wavenumbers between 600 and 4000 cm$^{-1}$ with a resolution of 4 cm$^{-1}$. A 15x objective was used with a slit aperture configured to acquire spectra over a rectangular region 25 \textmu m x 200 \textmu m, where the long dimension was parallel to the articular surface. Fifteen background-corrected scans were repeated at a given measurement point and averaged to generate a single IR spectrum. The acquisition window was scanned along the tissue sample’s depth at 25 \textmu m intervals by a computer-controlled stage to acquire measurements throughout the depth of the tissue. This procedure was repeated three times for each tissue sample with each scan separated laterally by roughly 2 mm taking care to avoid blood vessels and other artifacts.

We obtained the relative concentrations of aggrecan and collagen from the measured spectra by fitting the spectral window between 900 to 1725cm$^{-1}$ \cite{silverberg_structure-function_2014,didomenico_measurement_2019} for each measurement point with a linear combination of previously measured spectra for collagen and aggrecan \cite{camacho_ftir_2001}, as well as a linear background. This method is based on Beer-Lambert’s law, which states that IR absorbance is proportional to molecular concentration, and two mixed species of molecules have additive contributions. Including a linear background correction accounts for the instrument-specific deviations and drift which can occur in different labs for different environmental conditions \cite{smith_fundamentals_1995}. In fitting the spectra, we constrained the relative concentration of aggrecan and collagen to be greater than zero and assume that the linear background is the same throughout each sample. We did not consider the trace amounts of type IX collagen, type XI collagen, elastin, small nonaggregating proteoglycans, and other matrix macromolecules which occur in cartilage as their limited contributions to the absorption spectrum were negligible \cite{Luo_minor_2017}. The aggrecan and collagen concentration coefficients of the scans were then averaged for each depth, to determine the relative collagen and aggrecan concentrations as a function of depth.

\subsection{Mechanical Measurements} To prepare for mechanical testing, each hemi-cylindrical sample was soaked for 1h in PBS with 7 mg/mL 5-DTAF (5-dichlorotriazinylaminofluorescein) (Life Technologies, Carlsbad, CA), an all-protein stain. Samples were rinsed in PBS for 30 minutes to remove excess dye, then loaded into a Tissue Deformation Imaging Stage (Harrick Scientific, Pleasantville, NY). In this apparatus, the sample is gripped between two plates. Shear oscillations are applied to the surface of the sample, and the displacement of the second plate is used to measure shear stresses \cite{buckley_high-resolution_2010,buckley_localization_2013}. The sample was held in place between the shearing plates using cyanoacrylate glue to minimize the need for compression. Samples were immersed in PBS to maintain tissue hydration during mechanical testing.

The Tissue Deformation Imaging Stage (Harrick Scientific) was mounted onto an inverted LSM 5 Live confocal microscope (Carl Zeiss, Jena, Germany), where five lines were photobleached onto the rectangular surface of the hemi-cylinder perpendicular to the tissue surface. These lines caused no damage to the tissue and were used to facilitate automated computer tracking of the strain with a depthwise resolution of 10.4 \textmu m. Imaging of 1 Hz oscillatory shear at 1\% peak strain amplitude was carried out with a 10x objective, and movies of the tissue oscillations were acquired at 20 FPS throughout the entire depth of the tissue (Fig~\ref{fig:Method}). All mechanical testing was performed within 48 h of tissue harvest. The samples were retrieved after mechanical testing and frozen for biochemical assay.

Importantly, in order to isolate the effects of aggrecan on the tissue properties, our protocol entailed conducting our shear experiments under no compression. This procedure was necessary to avoid any effects due to buckling of the collagen network \cite{buckley_mapping_2008}, which would be much more localized in the degraded tissue. The main consequence of this procedure was that we observed much smaller changes in the moduli and energy dissipation for the degraded tissue than those observed in \cite{griffin_effects_2014,silverberg_structure-function_2014}. Avoiding this complication was also important for testing the rigidity percolation model since the model does not yet account for buckling of collagen fibers under compression. Since the normal function of cartilage tissue does entail compression, extending the model to address this regime would also be important.

Automated tracking of the local shear strain was facilitated by the five photobleached lines along the depth of the tissue. A custom Matlab code tracked the amplitude of the oscillations of the tissue, and the local shear strain was calculated by numerically differentiating the displacement with respect to the depth \cite{buckley_localization_2013}. The strain curves were fit to a sinusoidal function. The force applied to the tissue was able to be calculated by observing the displacement of the stationary shearing plate of a known stiffness. The depth-independent shear stress was able to be calculated by dividing this force by the surface cross section of the tissue. The shear modulus is this shear stress divided by the local shear strain. The difference in phase of oscillations between the local region and the applied stress are also recorded, which provides a measure of the viscous shear modulus relative to the total shear modulus.

\subsection{Biochemical Assay}

To determine the absolute concentrations of collagen and aggrecan in a tissue sample, we calibrated the coefficients obtained from the FTIR-I measurements using biochemical analysis \cite{Hoemann_Molecular_2004}. Mechanically tested hemi-cylinders were weighed for their wet weight, and frozen at -80$^\circ$C. The samples were dehydrated in a lyophilizer for 48 hours and then weighed again to obtain the dry weight. The samples were incubated in papain digest buffer (PDB) made from 125 mg/mL papain (Sigma) and 10 mM N-acetyl cysteine (Sigma) in PBD buffer (100 mM phosphate and 10 mM EDTA, pH 6.5) for 12 hours at 60 $^\circ$C. 

To measure the aggrecan concentration, GAG standards were created by dissolving chondroitin sulfate in PBD to known concentrations from 2 \textmu g/ml to 250 \textmu g/ml. 50 \textmu L of each standard and 50 \textmu L of the digest solution were added to a 96 well plate. 250 \textmu L of pH3 DMMB dye (Sigma) were added to each well. The plate was shaken for 30 seconds then the absorbance was read at 525 nm. The absorbance of the samples were matched to the absorbance of the known standards.

To measure the collagen concentration, hydroxyproline standards were created by dissolving hydroxyproline in PBD to known concentrations from 2 \textmu g/ml to 1000 \textmu g/ml. 50 \textmu L of 2M NaOH was added to 50 \textmu L of each standard and samples. The standards and samples were heated to 110 $^\circ$C for 18 hours. 30.5 \textmu L of HCl solution, 100 \textmu L 0.001 M CuSO$_4$, 100 \textmu L 2.5M NaOH, 100 \textmu L 6$\%$ H$_2$O$_2$, were each added to the standards and samples, ensuring to vortex after each addition. The standards and samples were allowed to rest at room temperature for 2 hours. The samples were vortexed and heated to 80 $^\circ$C for 5 minutes, then frozen. 400 \textmu L 3 M H$_2$SO$_4$ was added to each standard and sample, then they were frozen again. 200 \textmu L of DMAB (Sigma) were added to each standard and sample, then they were heated at 70 $^\circ$C for 15 minutes. 200 \textmu L of each standard and sample were added to a 96 well plate, and the absorbance was read at 540nm. The absorbance of the sample was matched to the absorbance of the known standards. Collagen is approximately 13.5$\%$ hydroxyproline, so the measure of hydroxyproline was multiplied by 7.4 to find the concentration of collagen in the sample. 

The total concentrations of collagen and chondroitin sulfate were taken from the biochemical assays and divided by the volumes of each sample, approximated by the depth of the sample measured during mechanical testing multiplied by its surface cross section. The concentrations were averaged for the healthy samples, and the degraded samples, and the FTIR-I curves were calibrated to the total average concentrations \cite{didomenico_measurement_2019}.

\subsection{Mathematical Model}

Rigidity percolation theories model biopolymer networks as disordered fiber networks and provide a framework to connect their rigidity to the network structure, composition, and single filament properties. These models have been immensely successful in predicting the mechanical properties and phase transitions of {\it {in vitro}} cytoskeletal and extracellular matrix networks as a function of filament concentrations. Previously, we combined this framework with a lattice-based  disordered fiber network reinforced by a background gel to explain the depth-dependent shear properties of AC\cite{silverberg_structure-function_2014}. In this framework, the shear modulus of the tissue was taken to be a sum of a background shear modulus, $\mu$, and a simulated shear modulus, $G_{sim}(\kappa/\alpha, \mu/\alpha, p)$, where $\kappa$ and $\alpha$ denote the bending and the stretching moduli of the collagen fibers, and  $p$ is the bond occupation probability of the collagen network, respectively.  Here we adapted this model to predict the mechanical response of both healthy and enzyme degraded articular cartilage. 

Since degradation with trypsin removes the aggrecan leaving the background hyaluronic acid network nearly intact, we extended the model to describe separate contributions of aggrecan and the associated hyaluronic acid network. Furthermore, the background modulus also includes mechanical reinforcement from collagen fibers that do not participate in the formation of a percolated network. Therefore, we expand $\mu$ about the bare HA value $\mu_0$, to linear order in aggrecan and collagen concentrations,  $\rho_a$ and  $\rho_c$, respectively, as $\mu= \mu_0 + \beta \rho_a + \gamma \rho_c$, where $\beta$ and $\gamma$ are expansion coefficients capturing the reinforcement of the bare hyaluronic acid network by aggrecan and collagen.  We computed the full shear modulus as 
\begin{equation}
 G = c \Big[\mu + G_{network}(\kappa/\alpha, \mu/\alpha, p)\Big],
\end{equation}
where $c$ is a scaling factor from simulation to experimental units, and the network modulus $G_{sim}$ was obtained from the simulations by minimizing the deformation energy density of the system, $\mathcal{E}$, under a given small strain $\epsilon_s$ and calculating $\frac{\partial^2 \mathcal{E}}{\partial {\epsilon_s}^2}$ (See SI for model construction and simulation details).

\subsection{Determination of Model Parameters}

We assumed the collagen fibers to be rods with a cross-sectional radius $r \sim 10$nm \cite{poole_immunoelectron_1982} and Young's modulus $E \sim 1$ GPa \cite{yadavalli_nanoscale_2010}. 
In the simulations, lengths and displacements were scaled by the bond length $l_c$, and rigidities were scaled by the fiber stretching modulus $\alpha$.  
This meant that for a fiber of length $l_c$, $\alpha \sim E r^2 /l_c$ and the scaled bending rigidity $\kappa \sim E r^4 /l_c^3$, the bending to stretch ratio was given by $\kappa/\alpha \sim r^2/l_c^2 \sim 10^{-4}$ for $l_c = 1 \mu$m. \cite{silverberg_structure-function_2014}.   
The elasticity of a densely connected network is stretching-dominated, and we expected the shear modulus to scale as $G \sim \frac{\alpha}{l_c} \sim \frac{E r^2}{l_c^2} \sim 10^5$ Pa,
setting the order of magnitude for $c$.  To inform an estimate of the stiffness for the HA gel, we used the data of Holmes, et al. ~\cite{holmes_hyaluronic_1988}, who found the concentration of 
hyaluronic acid in articular cartilage to be of the order of 1 mg/ml, with a  molecular mass on the order of 1 MDa. The classical estimate for a shear modulus of an aggregate of Gaussian chains, with a concentration of $c$ chains per unit  volume is $\sim k_{B} T c$ ~\cite{pethrick_polymer_2004}, yielding a shear modulus of the order of  1 KPa. These results suggested the normalized ratio of shear modulus to stretching
stiffness, $\mu_0 l_c / \alpha$ should be $\sim 10^{-2}$. As the product of $c$ and $\mu_0$ should yield a modulus of the order $10^3$ Pa, this separately suggests $c$ should be of order $10^5$.


We fitted for $c$, $\beta$, $\gamma$, $\mu_0$ and $\kappa$, using the above considerations to constrain our search. Parameters were chosen using $\chi^2$ minimization via the Nelder-Mead simplex algorithm in Mathematica. We found optimal values of $c = 2.33 \times 10^6$ Pa, $\mu_0 = .001$,  $\beta = 2.52 \times 10^{-3} $ ml/mg, $\gamma = 1.49 \times 10^{-4}$ ml/mg, $\kappa = .01$ and $\delta = .0052$, with $\chi^2 / D.O.F = 3.58$). The $\chi^2$ value for
each data bin was computed using the log base 10 of the experimental and model shear moduli.
See SI for further details.}

\showmatmethods{} 

\acknow{This work was supported by the National Science Foundation grant DMR-1807602, DMR-1808026,  CBET-1604712, CMMI 1927197, and BMMB-1536463. This work was also supported by the NIH National Institute of Arthritis and Musculoskeletal and Skin Diseases, Contract: 5R01AR071394-04. Finally, this work made use of the Cornell Center for Materials Research Facilities supported by the National Science Foundation under Award Number DMR-1719875.}

\showacknow{} 

\bibliography{main}

\end{document}


\bibliographystyle{apsrevM}

\title{Supplementary Information}

\date{\today}

\begin{abstract}
 \end{abstract}

\maketitle
\begin{figure}
\includegraphics[width=\columnwidth]{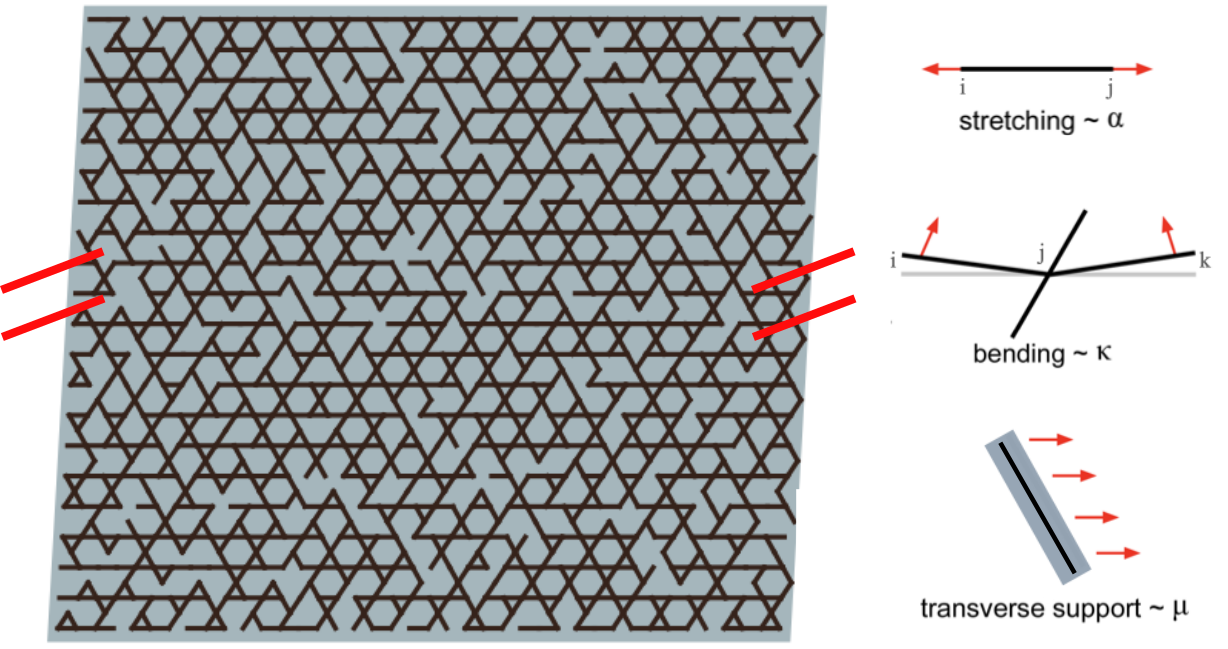}
\caption{{\bf Rigidity Percolation Model.} Shown is a schematic of a portion of the simulated kagome network. Hash marks indicate continuous boundary conditions in the lateral dimension. The black links represent the collagen fibers and the gray represents the background aggrecan gel. The links or bonds in the network are characterized by a stretching modulus $\alpha$ and a bending stiffness $\kappa$. The background gel further couples the links in the network through inhibition of transverse transport. This gel is characterized by a modulus $\mu$.}
\label{schematic}
\end{figure}

The model consists of a disordered kagome network made of stiff collagen fibers embedded in a continuum elastic background gel made of aggrecan and hyaluronic acid (Fig.\ref{schematic}), constructed following the procedure described in  
\citep{silverberg_structure-function_2014}.
The bonds in the network are randomly removed with probability $1 - p$, where $0 <p <1$, and a continuous series of colinear bonds constitute a fiber.  The fibers have a spring stiffness $\alpha$ and a bending modulus $\kappa$. The background gel resists fiber deformations in the transverse direction and its elasticity is described by the shear modulus $\mu$. The energy cost of deforming this network in the linear response regime is given by: 
\begin{eqnarray} 
\label{efunc}
E & = & 
\frac{\alpha}{2} \sum_{\langle ij \rangle} p_{ij}  \left (\textbf{u}_{ij} \cdot \hat{\textbf{r}}_{ij} \right)^2 \nonumber \\
&& + \frac{\kappa}{2} \sum_{\langle ijk \rangle} p_{ij} p_{jk}  \left[ \left( \textbf{u}_{ji} 
+ \textbf{u}_{jk} \right) \times \hat{\textbf{r}}_{ji}  \right]^2  \\ \nonumber
& & + \frac{\mu}{2} \sum_{\langle ij \rangle}   p_{ij} \left[ \textbf{u}_{ij}^2 -  \left( \textbf{u}_{ij} \cdot \hat{\textbf{r}}_{ij} \right)^2 \right],
\label{sim1}
\end{eqnarray}
where the first term corresponds to the energy cost of fiber stretching, the second term to fiber bending, and the third term to deformation of the background gel \citep{silverberg_structure-function_2014}.

The indices $i,j,k$ are summed over all nodes in the network, with $p_{ij}$ defined to be 1 when a bond between lattice sites $i$ and $j$ is present and 0 if such a bond is not present. The quantities $\hat{\textbf{r}}_{ij}$  and $\textbf{u}_{ij} = \textbf{u}_{i}-\textbf{u}_{j}$ are respectively the unit vector along bond $ij$ and the corresponding relative displacement. The bond length was set to unity in Eq.(\ref{sim1}), which renormalizes the units of distance used in calculating each elastic coefficient. Simulations were carried out for the parameter ranges  $10^{-4} <\kappa/\alpha  < 10^{-2}$, and $10^{-3} <\mu/\alpha< 1$. The applied compressive strain was $5\%$, while a range of shear strains centered about $1\%$ were applied to compute the linear response. We adopted a simple shear protocol in which external deformations were applied along the top boundary, vertices on the bottom boundary were fixed, and periodic boundary conditions were used for the left and right sides of the network. For each set of parameters, five networks containing $\sim 10^5$ nodes were randomly generated with a fraction $1-p$ of bonds missing, with $p$ varied from .4 to 1 in steps of .05.  The deformation energy (Eq.\ref{sim1}) was minimized for the applied  macroscopic compression and shear, and the shear modulus was calculated as a function of the bond occupation probability $p$ \citep{silverberg_structure-function_2014}.

To find an energy minimum, we note that the energy in \eqref{efunc} is
quadratic in the components of the $2N$--dimensional displacement field, 
allowing the energy to be written as a bilinear form with a stiffness matrix,
$\mathbf{K}$:

\begin{equation}
	E = \frac{1}{2} \sum_{l = 1}^{2N} \sum_{m=1}^{2N} K_{l, m} u_l u_m,
\end{equation}
and seek a zero force state, given by

\begin{equation}
    F = -\mathbf{K} \vec{u} + \vec{F}_C,
\end{equation}
where $\vec{F}_C$ is a vector of constraint forces needed to impose a purely affine
displacement field upon top and bottom nodes.

To isolate for nodes not on the top or bottom, here referred to as
interior nodes, we define a projection operator, $P_{N \rightarrow R}$ from the 
$2N$--dimensional space of all displacement field components to the
$2R$--dimensional space of displacements of the $R$ interior nodes. We also
define an operator $P_{R \rightarrow N}$ from the $2R$ to the $2N$--dimensional
space such that, for a full $2N$--dimensional displacement vector $\vec{u}$

\begin{equation}
	(P_{R \rightarrow N} P_{N \rightarrow R} \vec{u})_i =
	\begin{cases}
		u_i, i\hspace{1ex} \text{an interior coordinate}\\
		0, i\hspace{1ex} \text{a boundary coordinate}
	\end{cases}
\end{equation}

We next separate the displacement field into two components: an affine part,
$\vec{u}_A$, and a non-affine part, $\vec{u}_{NA}$, where, for a vertex with
initial coordinates $(x, y)$,

\begin{equation}
	\vec{u}_A(x, y) = (\epsilon_s y, -\epsilon_c y),
\end{equation}
where $\epsilon_s$ and $\epsilon_c$ are the shear and compressive strain,
respectively. For well-connected networks, we expect the displacement field
yielding an energetic ground state to very nearly equal the affine displacement
field, while for networks near the rigidity percolation threshold, large
departures are possible \cite{didonna05}.

We look for a non-affine displacement vector that leads to a zero overall force on all
interior nodes. Denoting by ${\vec{u}}_R$ the projection of the 
displacement field in the reduced, $2R$--dimensional subspace, we solve

\begin{equation}
	\label{force_eq}
	P_{N \rightarrow R} \mathbf{K} P_{R \rightarrow N} \vec{{u}}_{R, NA} = -P_{N \rightarrow R} \mathbf{K} \vec{u}_A,
\end{equation}
using the QR factorization method in SuiteSparse \cite{davis11}. Notably, this
solver can find pseudoinverse solutions even when the left-hand matrix is
rank deficient, and we verified upon solution that residual forces were of order
$10^{-12}$ or less. After obtaining a solution to equation ~\eqref{force_eq}, we
compute the full displacement field, $\vec{u}$, as

\begin{equation}
	\vec{u} = P_{R \rightarrow N} \vec{u}_{R, NA} + \vec{u}_A,
\end{equation}
and obtain the strain energy according to ~\eqref{efunc}.

We define the shear modulus as the second partial derivative of strain energy
with respect to shear strain, $\epsilon_s$, at fixed compressive strain 
$\epsilon_c = 5\%$:
\begin{equation}
	G =\frac{1}{A} \frac{\partial^2 E}{\partial \epsilon_s^2}\Bigg|_{\epsilon_c = 5 \%}.
\end{equation}
The second derivative was approximated using a centered finite difference 
approximation, with shear strains ranging from $.8\%$ to $1.2\%$, in steps of 
$\delta \epsilon = .1\%$, according to:

\begin{multline}
	\frac{\partial^2 E}{\partial \epsilon_s^2} \approx \frac{1}{12 \delta \epsilon^2} \Bigg[16\left[u(\epsilon_s^* - \delta\epsilon) + u(\epsilon_s^* + \delta\epsilon)\right] \\
	- u(\epsilon_s^* - 2\delta\epsilon) - u(\epsilon_s^* + 2\delta\epsilon) - 30u(\epsilon_s^*)\Bigg] + \mathcal{O}(\delta\epsilon^4),
\end{multline}
where $\epsilon_s^*$ denotes the target strain of $1\%$.

Experimental data were binned according to collagen concentration, with bin widths of $10 \hspace{2pt} \text{mg} / \text{ml}$. Within each bin, we found the arithmetic mean of collagen concentration and aggrecan concentration, and the  geometric mean of shear modulus. The uncertainty in the shear modulus was  estimated using the geometric standard deviation.

From discrete simulation data, we constructed a fourth-order interpolation to estimate the shear modulus of the tissue, which was taken to be a sum of a background shear
modulus, $\mu$, and a simulated shear modulus, $G_{sim}(p, \kappa, \mu)$, where $\kappa$ and $p$ denote bond bending stiffness and bond occupation probability,
respectively. The shear modulus, $\mu$, is presumed to have a base value of $\mu_0$. We expand $\mu$ to linear order in both aggrecan concentration, 
$\rho_a$, and collagen concentration, $\rho_c$. We compute the full shear modulus as

\begin{equation}
	G = c \left[ (\mu_0 + \beta \rho_a + \gamma \rho_c) + G_{sim}(\delta \rho_c, \kappa, \mu_0 + \beta \rho_a + \gamma \rho_c)\right]
\end{equation}
Here, $c$ is a scaling factor from simulation to experimental units, $\beta$ and $\gamma$ are expansion coefficients capturing the reinforcement of
the bare hyaluronic acid gel by proteoglycans and collagen, and $\delta$ relates collagen concentration to bond occupation probability. We model $p$ as linearly dependent upon collagen concentration, such that the maximum experimentally observed collagen concentration corresponds with $p = 1$.

The $\chi^2$ statistic is computed by comparing the $\log_{10}$ of the geometric mean of the shear modulus for each bin to the model value predicted
using the arithmetic means of collagen and aggrecan concentrations. We optimize fitting parameters by minimizing

\begin{equation}
        \chi^2 = \sum_{i = 1}^N \frac{\left[ \log_{10}(\bar{G}_{exp, i}) - \log_{10}(G_{model}(\bar{\rho}_{c,i}, \bar{\rho}_{a,i}))\right]^2}{\log_{10}(\sigma_i)^2},
\end{equation}
where $N$ is the number of bins, $\bar{G}_{exp, i}$ is the geometric mean of the shear modulus for the $i$th bin, $\bar{\rho}_{c,i}$ and $\bar{\rho}_{a,i}$
are the arithmetic means of collagen and aggrecan concentrations for the $i$th bin, and $\sigma_i$ is the geometric standard deviation of the shear modulus in
the $i$th bin.

We fit our model to degraded data to obtain values for $c$, $\mu_0$, $\beta$ $\gamma$,
and $\kappa$, while setting the value of the $\delta$ to be the reciprocal of
the maximum collagen concentration encountered in experiment. Parameters were 
constrained to be within physiologically and experimentally relevant ranges,  
and were optimized using $\chi^2$ minimization via the Nelder-Mead simplex 
algorithm in Mathematica. We obtained the value of $\beta$ by fitting the
model to healthy data, while using the values obtained from degraded data for
all other parameters. We found optimal values of $c =  2.33 \times 10^6$ Pa,
$\mu_0 = .001$,  $\beta = .0025 $ ml / mg, $\gamma = 1.49 \times 10^{-4}$ ml / mg,
$\kappa = .01$ and $\delta = .0052$, with $\chi^2 / D.O.F = 3.58$. We tested
for possible degradation of collagen by attempting to fit for a separate 
bending rigidity of collagen for healthy tissue, and found a negligible change.


We finally considered the possibility of treating the network and gel entirely
separately, so that the third sum in ~\eqref{efunc} is eliminated, and the
shear modulus is modeled according to

\begin{equation}
	G = c \left[ \mu_0 + \beta \rho_a + \gamma \rho_c + G_{sim}\left(\delta \rho_c, \kappa\right) \right].
\end{equation}
We optimized parameters as described above, finding $c = 2.37 \times 10^6 Pa$,
$\mu_0 = 10^{-3}$, $\beta = 5.07 \times 10^{-4}$ ml / mg, 
$\gamma = 3.58 \times 10^{-4} ml / mg$, $\kappa = 10^{-2}$, and 
$\delta = 5.2 \times 10^{-3}$, with $\chi^2 / D.O.F. = 6.7$. This model 
exhibited considerably worse agreement with experimental observations than
the approach that considers direct coupling between the network and background
gel. In particular, ignoring direct coupling leads
to predictions for the shear modulus that are
systematically too high for low collagen concentration. This suggests that ignoring the
interaction between the collagen scaffold and the
aggrecan matrix eliminates an important means of
reinforcing cartilage, leading to an overprediction
of the shear modulus of aggrecan to compensate for
this deficit. We thus fund the coupling term in
the third sum of ~\eqref{efunc} to significantly
improve the predictive power of our model.

\bibliography{main}